\documentclass[12pt]{iopart}
\usepackage{amssymb}
\usepackage{adjustbox}
\usepackage[T1]{fontenc}
\usepackage[utf8]{inputenc}
\usepackage{dirtytalk}
\usepackage{xcolor}		
\usepackage{soul}		
\usepackage{iopams}  
\begin{document}

\title[Low-cost portable turbidimeter]{Development of a low-cost portable turbidimeter for processes}

\author{L C Sperandio$^1$, M S Colombo$^1$, 
C M G Andrade$^1$\footnote{Present address:
Department of Chemical Engineering, 
State University of Maringa, Colombo av., Maringa, BR} and C B B Costa$^1$}

\address{$^1$ Department of Chemical Engineering, 
State University of Maringa, Colombo av., Maringa, BR}
\ead{cidmga@yahoo.com.br}
\begin{indented}
\item[]April 2021
\end{indented}
\vspace{10pt}

\begin{abstract}
Turbidity is a physical property related to the scattering of light by particles that are suspended in a liquid. Commercial turbidimeters are priced at the range of hundreds to thousands of dollars. Considering this scenario, it is proposed in this work the development of a low-cost portable turbidimeter for monitoring of turbidity in processes. An infrared LED was used as light emitter, and an infrared phototransistor as light receiver. The signal processing control unit was developed with the Arduino Uno platform. The calibration of the turbidimeter was done by means of a comparative test in triplicate, using as reference the commercial turbidimeter 2100P, HACH®. The turbidimeter was able to perform analysis in the range of 100 to 1000 NTU, presenting an innovation character given its portability and computer communication via USB, and in a good price range for the prototype, costing US\$ $46.30$.
\end{abstract}

%
\vspace{2pc}

%
\submitto{\JINST}
%
%
%

\section{Introduction}

Turbidity monitoring is an important method of quality control, widely used in many food, chemical, pharmaceutical and processing industries \cite{metzger2018low}. The analyzes of this property can be applied in the quality monitoring of a variety of fluids, ranging from the raw materials to the resulting processed goods. An example is the application in the production of crystal sugar, where the turbidity of the sugarcane juice is constantly monitored to guarantee the conditions of clarity of the juice \cite{Rodrigues2018}. It can also be used for monitoring effluents, by the industry itself or by governing agencies inspecting fluid discharges into open waters.

Turbidity is a physical property related to the presence of suspended particles, leading to loss of clarity of the liquid. In physical concepts, it is related to the intensity of light that is scattered as it propagates in the liquid when interacting with the suspended particles and therefore being deflected in different directions. The higher the amount of suspended particles, the greater the turbidity of the liquid. The most commonly used turbidity unit is the nephelometric turbidity unit \cite{sampedro2015turbidimeter}.

Modern turbidimeters are based on the transmission and scattering of light from a source, such as a light emitting diode (LED). Typically, photodiodes or phototransistors are used as light receptors. This radiation emitters and receptors can be implemented in various geometrical arrangements. There can be also a number of extra receptors, in different positions related to the path of light, with the objective of improving accuracy of measurements in either extremes of suspended particles concentration. Either low or high concentration of samples are difficult to accurately measure the turbidity.

The single beam turbidimeter is the simplest modern model available. It consists of one light emitter and one light detector, and can be projected from two measurement techniques: turbidimetry and nephelometry. In turbidimetry, the detector and the emitter are positioned with their optical axes on the same line, i.e. facing each other. In this way, the reduction of light intensity through the liquid to be measured is detected \cite{Fetisov2009}. The turbidity technique works better for measurement of samples with high concentration of suspended particles.

\begin{figure}[h]
\centering
\includegraphics[width=0.85\linewidth]{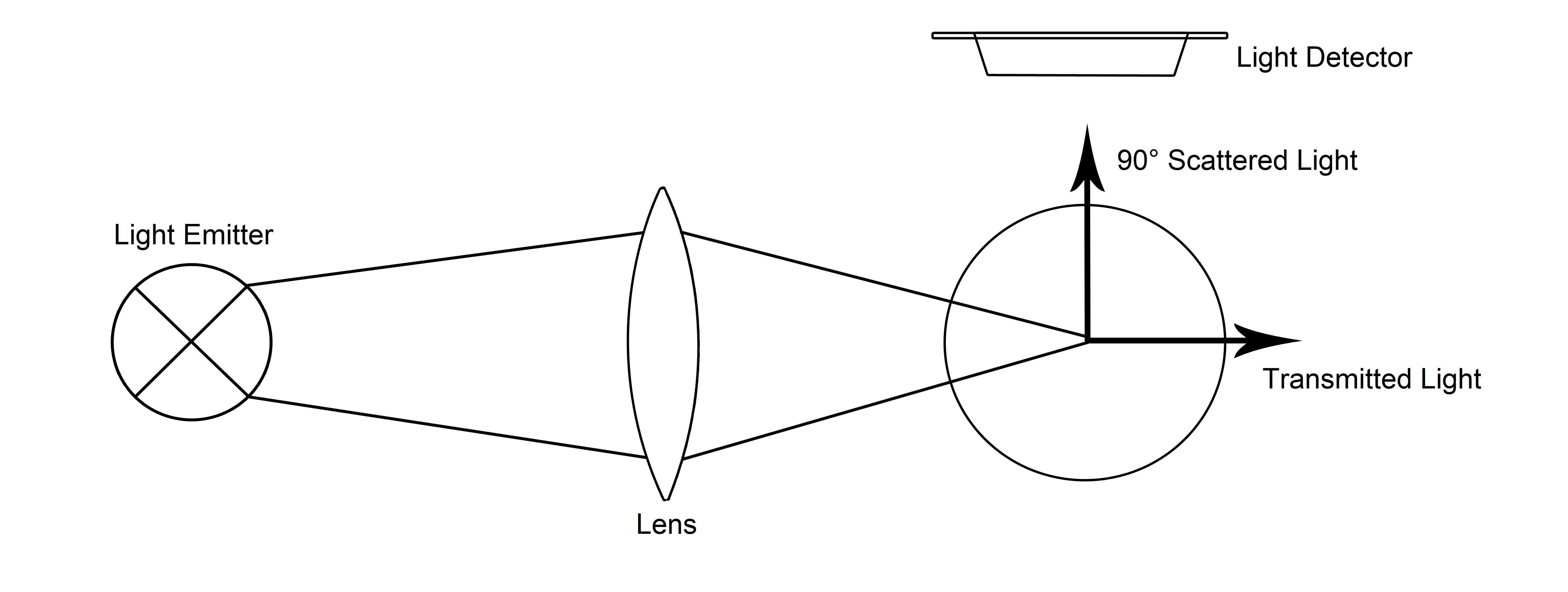}
\caption{Single beam model using the nephelometry technique (detection of light scattered at 90°}
\label{fig1}
\end{figure}

In nephelometry, the detector is positioned at an angle of 90° to the light emitter to capture the light that has been scattered. The greater the intensity of the scattered light detected, the greater the turbidity of the liquid \cite{khairi2015review}. The operating principle of a turbidimeter using the nephelometry technique is illustrated in figure \ref{fig1}.

Many commercial turbidimeters are based on the nephelometry principle, with prices ranging from a few hundred to several thousand US dollars. In this scenario, the development of a low-cost device that is able to measure the turbidity of a sample in a curvette holder becomes a subject of interest \cite{kelley2014affordable,RomanHerrera2016,metzger2018low}. Aisopou, Stoianov, and Graham \cite{Aisopou2012} developed a single beam turbidimeter, with scattered light detection, to be installed directly in a water distribution line. The paper of Omar and MatJafri \cite{Omar2012} introduces a significant work by comparing different low cost configurations using near infrared sensors, using fiber optics or high sensitivity sensors. Wiranto, Hermida, and Fatah in \cite{wiranto2016design} developed a single beam turbidimeter with a probe, using laser as a beam of light and the nephelometry technique.

Over the last few years, open source technologies (or free technologies) have gained a lot of space in the development of software and products. The use of hardware or software with open source philosophy in turbidimeters is not a novel approach \cite{kelley2014affordable,wiranto2016design,RomanHerrera2016,metzger2018low,nguyen2018low}.   In the field literature, the motive for developing such low cost devices is to provide, for third world countries, an alternative to the costly commercial turbidimeters available, applied to potable and drinking water analyzes. There is also a number of papers that describe the development of turbidimeters by making use of flash light source and the camera or the computational resources of smartphones \cite{Hussain2016,hussain2017low,Bayram2018,koydemir2019smartphone}. These solutions are low cost and can be seen as practical implementations for potable water quality tests.

Kelley \textit{et. al.} \cite{kelley2014affordable} proposed a low-cost turbidimeter to be used in low-income communities for drinking and potable water quality assessment. Their design utilizes cheap and simple parts, for instance a LED and light to frequency sensor, as well as a arduino type microprocessor. Another relevant aspect in that work was the calibration of their prototype with a colloidal suspension of oil in distilled water, instead of use of formazin as recommended by standards. This approach was also followed in the reference \cite{Kovacic2019}. One of these standards is from the Environmental Protection Agency (EPA 180.1), and the other is from the International Standards Organization (ISO 7027). Both the referenced standards recommend the use of formazin polymer for calibration of turbidity measurement hardware. Formazin is produced through a polymerization using hydrazine sulfate, a carcinogenic chemical compound. There are other several issues to be considered in applying formazin suspensions for turbidimeter calibration, for instance, stability of the suspension. \cite{munzberg2017limitations}.

As discussed in the review paper in \cite{Omar2009} by Omar and MatJafri, there was considered at the time serious limitations in the application of fiber optics for turbidity measurements. Fiber optics offers the advantage of separating sensors from the liquid sample, but it was argued by the authors that problems could arise from this configurations likewise resolution of the reading and ambient light pollution of the measurement. Later, there were several papers that successfully applied optical fiber to turbidimeters \cite{Omar2012,Arifin2017,nguyen2018low}.

Metzger \textit{et. al.} \cite{metzger2018low} developed a new nephelometric turbidimeter with an incorporated Graded Index Lens to separate the electronic components and the light source from being close to the sample vial. This strategy is also adopted in \cite{Bayram2018}, as they used fiber optics to channel transmitted light in the sample (absorbance phenomena) to the receptor sensor.

Nguyen and Rittmann \cite{nguyen2018low} reported an open-source arduino-based turbidimeter with an Infrared sensor, model TSD-10, used in washing machines. Kirkey, Bonner and Fuller \cite{kirkey2018low} used a RGB light with ambient light rejection mechanism for construction of a submersible turbidimeter. The resulting device was designed to be used in line, submersed directly in the body of water. Gillett, Marchiori \cite{Gillett2019} introduced a continuously turbidimeter to be installed directly in the current streams line, using a light to frequency sensor, LED infrared, and arduino-based system architecture.

Román-Herrera \textit{et. al.} \cite{RomanHerrera2016} also worked on the development of low-cost turbidimeters, using two white light LEDs as emitters and two light dependent resistors (LDR) as receivers. Kovacic and Asperger \cite{Kovacic2019} developed a turbidimeter capable of, acording the authors, measuring turbidity, colorimetry and nephelometry, calibrated with a formazin polymer suspension.

In view of the above, the objective of this work was the development of a low-cost portable turbidimeter (with the possibility of conversion to an online and inline turbidimeter), using the nephelometry technique and open source technologies, with applicability in turbidity monitoring as a parameter of process quality. As differentials, this device has a very low-cost, is portable and can communicate with a computer via the USB port.

\section{Materials and Methods}
\label{sec2}

In order to meet the defined objectives, a low-cost portable turbidimeter for liquid analysis was developed with great applicability in food processing industries, but also in process plants in general where it may be of interest to analyse turbidity of a liquid stream with the exception of potable water. The operating principle of the developed turbidimeter is the emission of an infrared light beam in the direction of the sample and the detection of the light intensity that is scattered at $90^{\circ}$ of the incident light beam (nephelometry). For the development of the turbidimeter, different items needed to be developed and/or implemented: measurement chamber, infrared emitter driver circuit, infrared receiver reading circuit, control and processing unit, peripheral components.

Because it is a low-cost turbidimeter, designing and manufacturing a cuvette has proved completely impractical. Using cuvette of a turbidimeter already existing in the market would be more viable, but still financially out of the proposal. Then, a search for an alternative sample container was performed, which was easy to acquire and inexpensive. The most interesting alternative, and that was adopted, was to use a blood collection tube.

The designed measurement chamber has been manufactured in a 3D printer, a device widely used today in design and prototyping. Acrylonitrile butadiene styrene (ABS) was used as raw material. The measurement chamber is a tube with one open end and the other closed, where the inner diameter being slightly larger than the outer diameter of the sample container, so that it can be inserted into the chamber. It has four holes, a design that allows further reasearch with the developed prototype using 2 emitters and 2 sensors as following the modulated four-beam technique, documented by Postolache and collaborators in \cite{postolache2007multibeam}. In the prototype and the results generated for this paper it was inserted in the measurement chamber the emitter and receiver, positioned at the same height and around the tube, equally separated from their adjacent ones by $90^{\circ}$, as required by the nephelometry technique. The measurement chamber is shown in Figure \ref{fig2}.

\begin{adjustbox}{center,caption={Measurement chamber developed in 3D printer},label={fig2},nofloat=figure,vspace=\bigskipamount}
\includegraphics[width=0.45\linewidth]{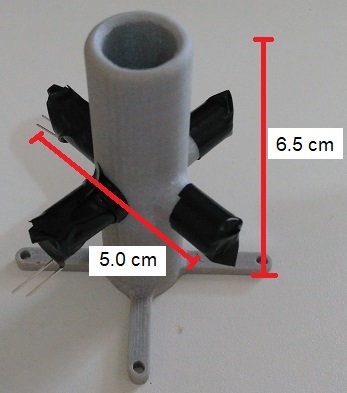}
\end{adjustbox}

To act as a light source in the turbidimeter, the PHIV459 LED was chosen. It emits infrared light with emission angle of $30^{\circ}$ and nominal wavelength of 940 nm when operating with electric current of 100 mA and voltage of 1.7 V. After defined the LED to be used as infrared emitter, its driver circuit was developed. This circuit must be able to receive the digital signal to drive the emitter from the D8 digital port of the control and processing unit, and supply it with a constant current of 50 mA. For this type of power, a current source known as current drain was chosen. The diagram of the LED driver circuit is shown in Figure \ref{fig3ab}.

As to the light receiver, as suggested by the manufacturers, it is typically used the pair of the employed emitter. In this case, it is the phototransistor PHFT458. This phototransistor has an angle of incidence of $ 25^{\circ}$ and a peak wavelength of 880 nm. The encapsulation and dimensions are the same as the LED PHIV459.

In order to perform the measurement, the phototransistor was used in a circuit known as the common emitter amplifier, operating in the active region. In this configuration, the measurement is performed at the collector and the emitter is grounded, with the result that the electrical voltage on the collector is inversely proportional to the infrared light intensity incident on the phototransistor. This voltage value should be between 0 V and 5 V, being read by A0 analog port of the control and processing unit and converted to a dimensionless value between 0 and 1023. The diagram of the phototransistor reading circuit is shown in Figure \ref{fig3ab}.

\begin{adjustbox}{center,caption={LED driver circuit (a) and phototransistor measurement circuit (b)},label={fig3ab},nofloat=figure,vspace=\bigskipamount}
\includegraphics[width=0.75\linewidth]{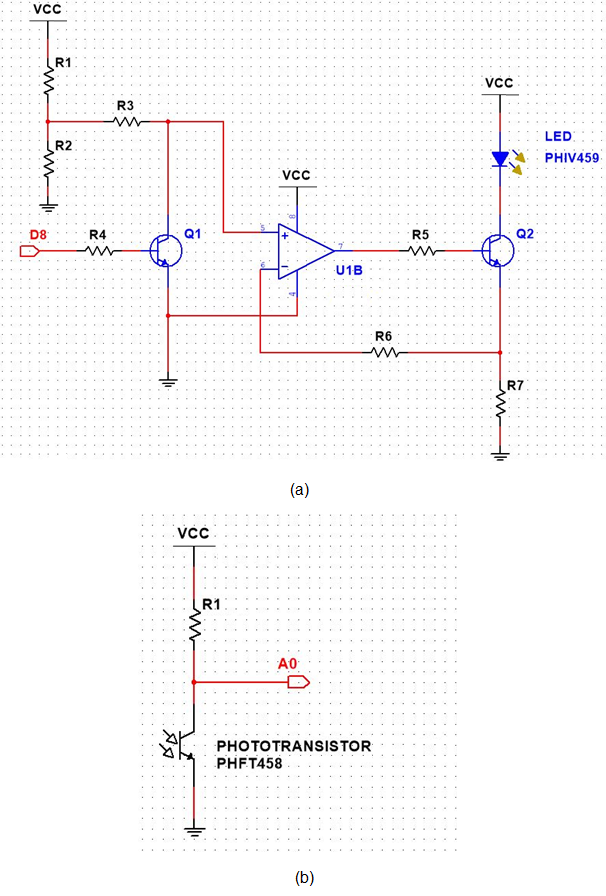}
\end{adjustbox}

Regarding the control and processing unit of the turbidimeter, the microcontrolled platform Arduino Uno was used. It is an open source and low-cost electronic prototyping platform, capable of receiving and processing inputs from sensors or other devices and then generating outputs for actuators or other devices, according to the programming developed by the user.

To power the entire system, a 9 V battery connected to the P4 connector of the Arduino was employed, which is suitable for external power (power without using the USB port). To turn the entire system on and off, a small on/off switch was used between the positive pole of the battery and the positive pole of the P4 connector, so that the switch could interrupt this connection when it was turned off, thus turning off the entire system.

As for the peripherals, a liquid crystal display of 16 x 2 characters and a push button were used. The screen has the function of instructing the operator when turning on the turbidimeter and then informing the measured turbidity value of the sample. When the button is pressed, connected to D9 digital port of the control and processing unit, the sample turbidity measurement is started.

With the development of each part of the turbidimeter completed, the next step was to assemble the device. Firstly, the electronic board was built by integrating the LED driver circuit and the phototransistor reading circuit. To facilitate and organize the connections, the connectors for power the liquid crystal display and the start button were integrated into the board. The resulting electronic board is shown in Figure \ref{fig4}.

\begin{adjustbox}{center,caption={Electronic board for LED driver and phototransistor measurement},label={fig4},nofloat=figure,vspace=\bigskipamount}
\includegraphics[width=0.50\linewidth]{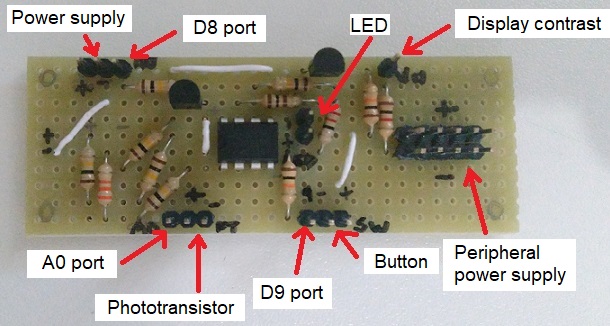}
\end{adjustbox}

Subsequently, all components were fixed to the turbidimeter structure. A plastic box with dimensions of 15 x 10 x 5.5 (cm) was used as structure. Once the turbidimeter assembly was completed, the complete system programming was developed. This programming is done on a computer, using Arduino Integrated Development Environment (IDE), and transferred to the Arduino Uno through the USB connection.

The device works as follows. The user turns on the turbidimeter with the on/off switch and waits until the message \say{INSERT SAMPLE AND PRESS THE BUTTON} appears on the screen. Then the sample container with the liquid to be analyzed should be placed in the measurement chamber and the button should be pressed to start the analysis. At the end of the measurement and processing steps, the dimensionless reading value and calculated turbidity value in NTU are displayed on the screen. If the turbidimeter is connected to a computer through the USB port, then the calculated turbidity value will also be sent through serial communication. After five seconds, the system returns to the beginning, being able to analyze a new sample.

In order to determine the calibration equation of the turbidimeter, which was responsible for converting the dimensionless value of the phototransistor reading into a turbidity value, given in NTU, a comparative method was used with reference to a commercial turbidimeter. The turbidimeter used is the 2100P model, manufactured by HACH®. The 2100P turbidimeter operates in the range of 0 to 1000 NTU, with accuracy of $ \pm  2 \%$ of reading. Thus, the developed portable turbidimeter could only be evaluated in this measuring range. Figure \ref{fig5} shows the developed turbidimeter and 2100P turbidimeter side by side in order to compare the physical dimensions.

\begin{adjustbox}{center,caption={Turbidimeter developed (left) and 2100P turbidimeter (right)},label={fig5},nofloat=figure,vspace=\bigskipamount}
\includegraphics[width=0.80\linewidth]{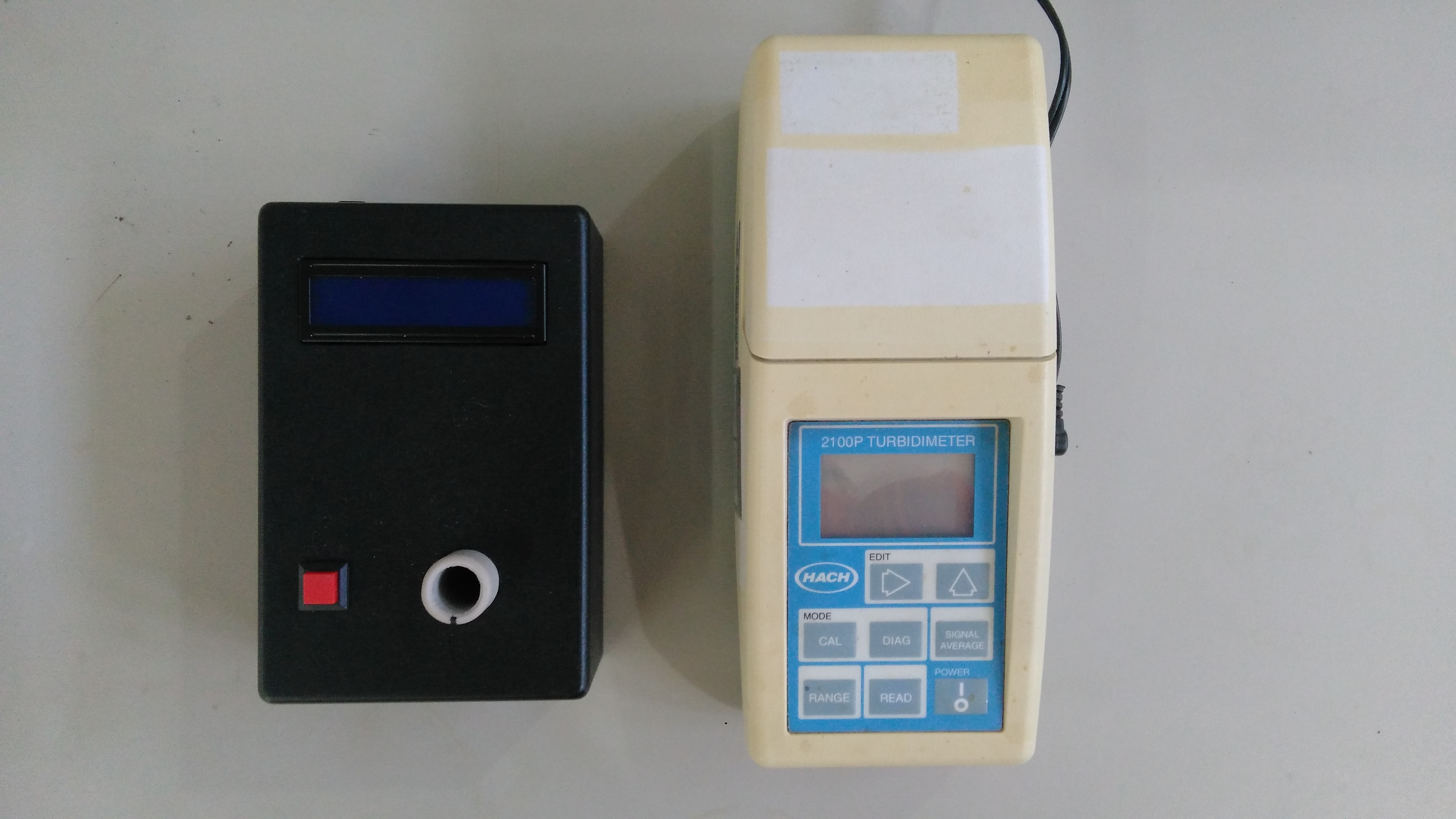}
\end{adjustbox}

Comparative tests were carried out using concentrated passion fruit juice, diluted in different proportions with deionized water obtained in a reverse osmoze water purifier system. As the turbidimeter was developed for applications in processes, for example, juice production, then the use of concentrated juice in this analysis is appropriate and convenient.

The calibration method consisted in producing a sequence of distinct samples and performing in each sample the turbidity measurement with the 2100P turbidimeter and the dimensionless reading in the developed turbidimeter, recording these two pieces of information together with the proportion of juice and water used to produce each sample. Three test sequences were performed in order to perform the study in triplicate. The calibration equation was obtained through regression, using the least squares method to determine the parameters.

\section{Results and Discussion}
\label{sec3}

Data collected in the three test sequences are presented in Table \ref{tabela1}, and their respective scatter plots and regressions are shown in Figure \ref{gra6}.

\begin{adjustbox}{center,caption={Graphs with all data referring to comparative tests},label={gra6},nofloat=figure,vspace=\bigskipamount}
\includegraphics[width=0.90\linewidth]{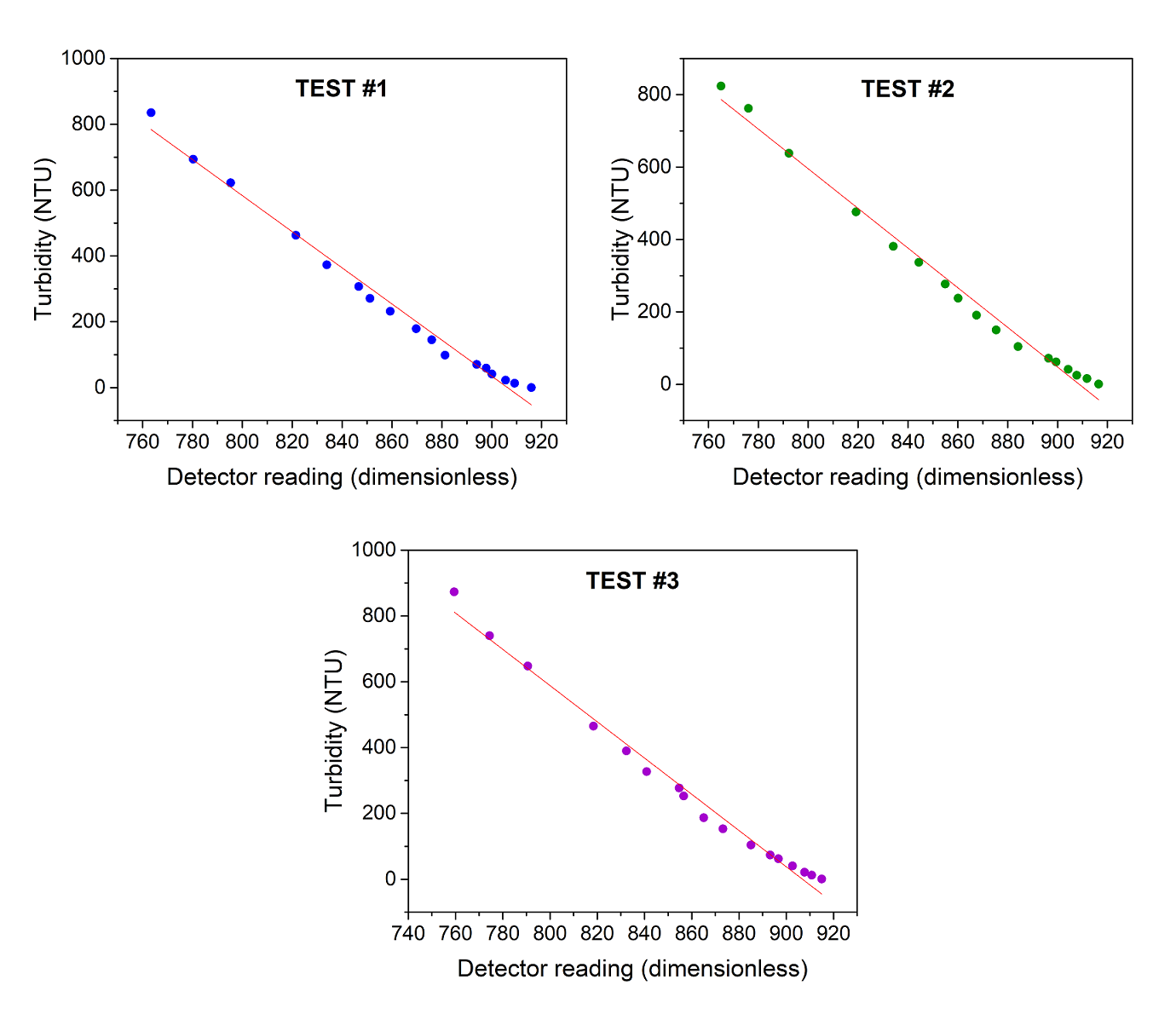}
\end{adjustbox}

\begin{table}
\caption{\label{tabela1} Test data in the comparative tests between the commercial turbidimeter 2100P and the low-cost portable turbidimeter developed for this work}
\footnotesize
\begin{tabular}{@{}lllllll}
\br
  & \multicolumn{2}{c}{Test \#1} &\multicolumn{2}{c}{Test \#2}& \multicolumn{2}{c}{Test \#3} \\ 
Sample  & 2100P  & Prototype  & 2100P  & Prototype  & 2100P  & Prototype  \\ 
(juice:water ratio) & (NTU) & (dimensionless)  & (NTU) &(dimensionless) & (NTU) & (dimensionless) \\ 
\mr 
5:70 & 835.00 & 763.4 & 824.00 & 765.1 & 873.00 & 759.5 \\ 
\hline 
5:80 & 694.00 & 780.3 & 762.00 & 776.1 & 740.00 & 774.4 \\ 
\hline 
5:90 & 622.00 & 795.3 & 638.00 & 792.3 & 648.00 & 790.6 \\ 
\hline 
5:115 & 463.00 & 821.5 & 476.00 & 819.2 & 465.00 & 818.4 \\ 
\hline 
5:135 & 373.00 & 833.9 & 381.00 & 834.1 & 390.00 & 832.4 \\ 
\hline 
5:155 & 307.00 & 846.7 & 337.00 & 844.4 & 327.00 & 840.9 \\ 
\hline 
5:175 & 271.00 & 851.2 & 277.00 & 855.0 & 277.00 & 854.7 \\ 
\hline 
5:195 & 232.00 & 859.3 & 238.00 & 860.1 & 253.00 & 856.6 \\ 
\hline 
5:235 & 179.00 & 869.7 & 191.00 & 867.5 & 187.00 & 865.1 \\ 
\hline 
5:275 & 145.00 & 876.0 & 150.00 & 875.4 & 153.00 & 873.2 \\ 
\hline
5:355 & 98.20 & 881.3 & 104.00 & 884.2 & 104.00 & 885.1 \\ 
\hline 
5:475 & 70.30 & 894.0 & 72.10 & 896.4 & 73.60 & 893.2 \\ 
\hline 
5:555 & 59.50 & 897.8 & 61.50 & 899.4 & 62.20 & 896.7 \\ 
\hline 
5:795 & 41.00 & 900.1 & 41.60 & 904.3 & 40.30 & 902.7 \\ 
\hline 
5:1115 & 22.80 & 905.6 & 24.90 & 907.7 & 21.30 & 907.8 \\ 
\hline 
5:1915 & 13.10 & 909.2 & 16.10 & 911.8 & 12.50 & 910.8 \\ 
\hline 
Pure Water & 0.13 & 915.9 & 0.45 & 916.5& 0.60 & 915.0 \\
\br
\end{tabular}\\
\end{table}
\normalsize

By analyzing more carefully the behavior of the data in the graphs of the three tests (Figure 6), it is possible to verify that in the samples with turbidity less than 100 NTU, the relation between the reading of the receiver and the value of the turbidity ceases to behave in a linear way. Thus, it is not appropriate to represent this range through this linear equation resulting from the regression. In addition, it can be seen that this non-linear tendency of the data of samples with turbidity less than 100 NTU ends up displacing the equation resulting from the regression, damaging the correct representation of the collected data that present linear behavior, which is the desired behavior in the turbidity measurement.

This behavior may be the result of an unsatisfactory resolution of the developed turbidimeter for the 0 to 100 NTU range. As the developed turbidimeter is applicable in industrial processes where high turbidity liquids are common, it was decided to exclude the data for the range of 0 to 100 NTU of the analysis and to determine the calibration equation for the range of 100 to 1000 NTU. The corrected scatter plots and the new regressions are shown in Figure \ref{gra7}.

\begin{adjustbox}{center,caption={Graphs with only used data referring to comparative tests},label={gra7},nofloat=figure,vspace=\bigskipamount}
\includegraphics[width=0.90\linewidth]{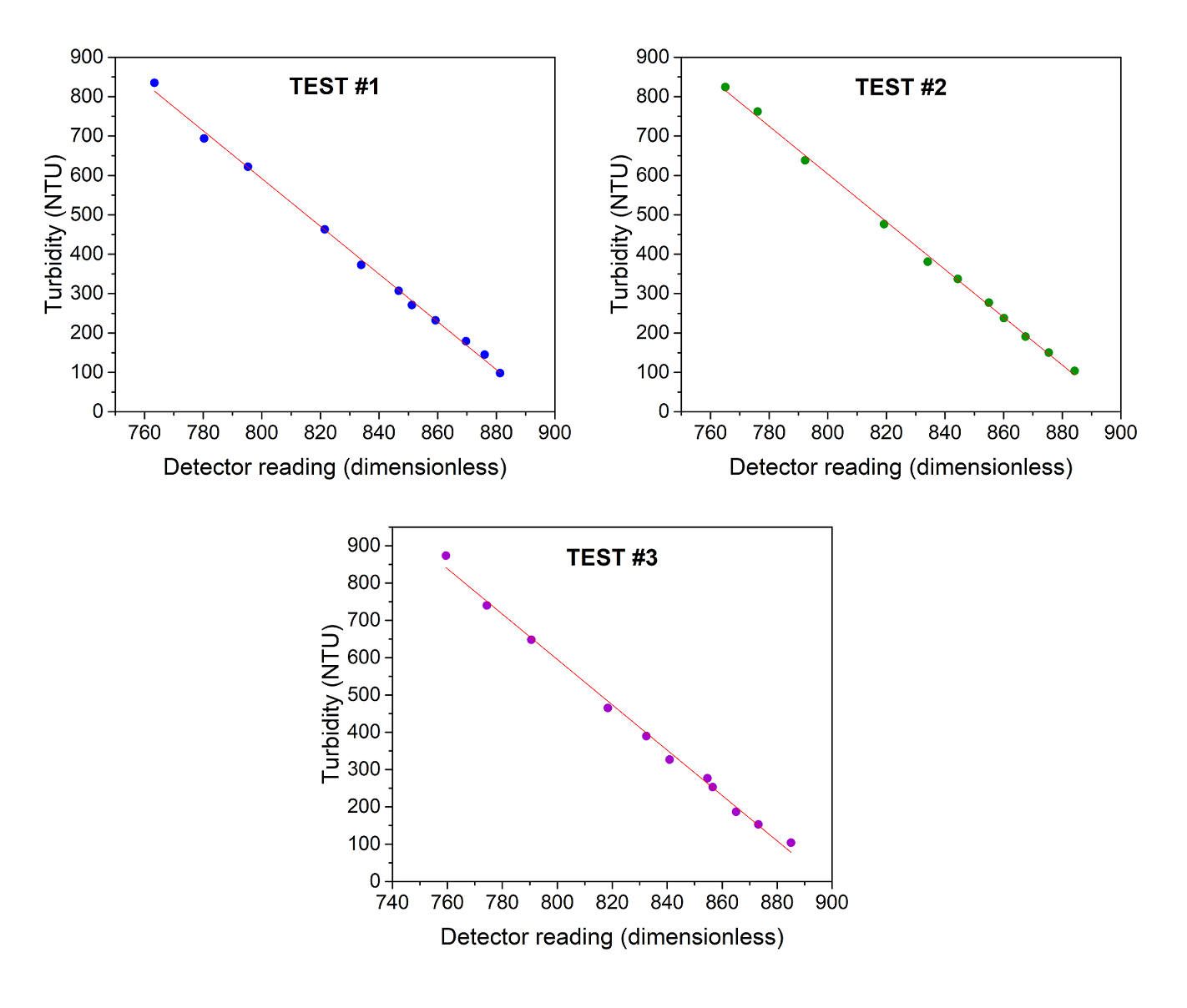}
\end{adjustbox}

Analyzing the graphs in the new working range, shown in Figure 7, it is evident that the data collected in the three tests, for the working range of 100 to 1000 NTU, showed the desired purely linear behavior. This is corroborated by the values of the respective adjusted coefficients of determination $R^2$, which were between $0.99$ and $1.00$. Thus, there is excellent representativeness of the data through the respective calibration equations. In addition, the calibration equations resulting from the regression of the data of each test have parameters of values very close to each other, characterizing good repeatability of the developed turbidimeter. In the three equations, the linear coefficients presented an error within the range of $\pm 67.68$ to $\pm 115.17$, while the angular coefficients presented error within the range of $\pm 0.08$ to $± 0.14$.

As only one calibration equation is required, it was chosen to generate an equation that is the result of the three tests performed. For this, the data of the three tests were grouped into a single analysis, resulting in a new graph, a new regression, and then a calibration equation that represents all the data collected. This new analysis is presented in Figure \ref{gra8}.

\begin{adjustbox}{center,caption={Graph combining data from the comparative tests},label={gra8},nofloat=figure,vspace=\bigskipamount}
\includegraphics[width=0.60\linewidth]{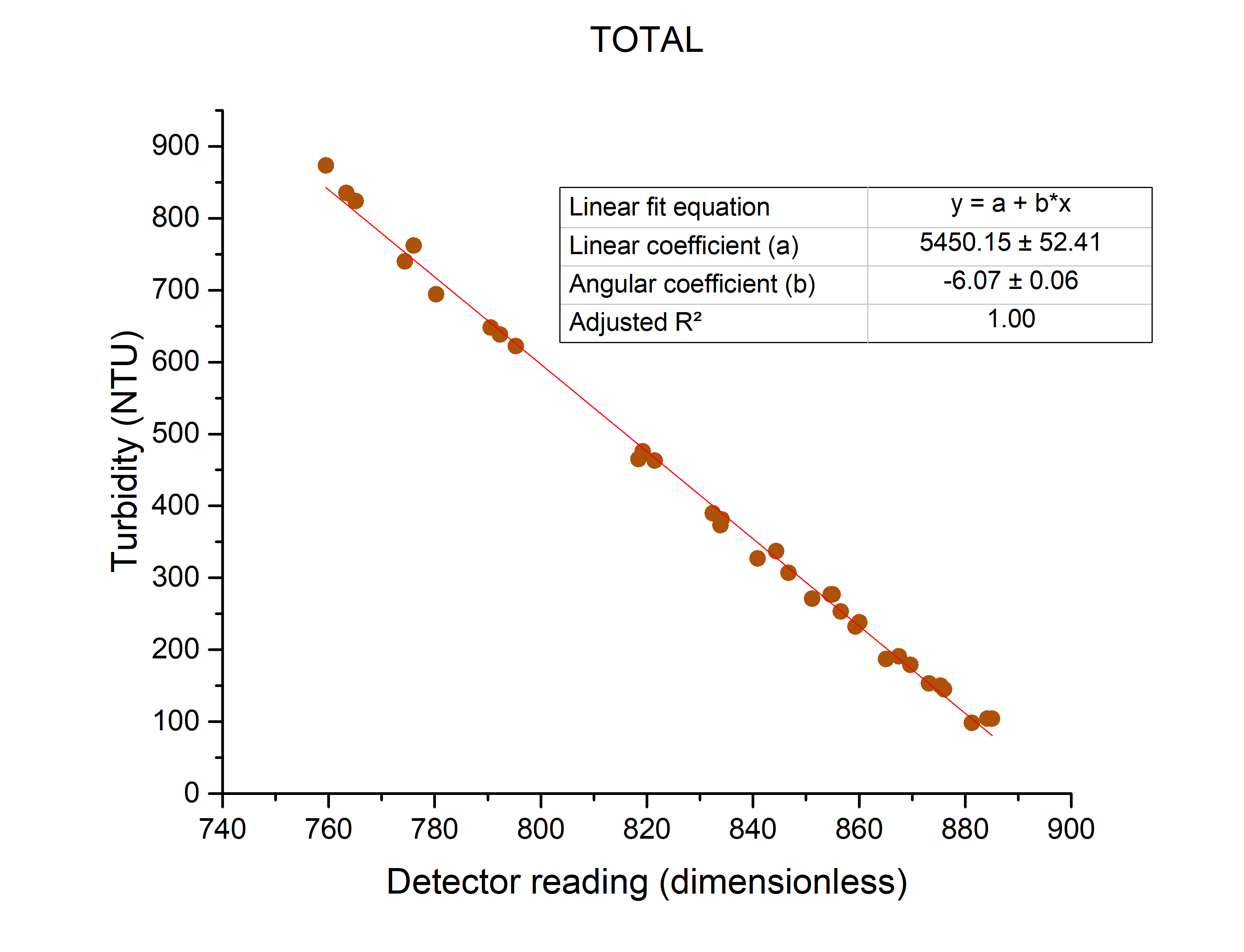}
\end{adjustbox}

Again, the data presented the purely linear behavior desired, resulting in an excellent quality regression with a coefficient of determination R² adjusted of $1.00$. The errors of the linear and angular coefficients of this regression are the lowest ones found until then, being $ \pm 52.41$ and $ \pm 0.06$, respectively. The resulting calibration equation is expressed in Equation \ref{eq4}, where x is the value of reading detector, dimensionless, and y is the value of turbidity, in NTU.

\begin{equation}
y=-6.07x+5450.15
\label{eq4}
\end{equation}

A relevent discussion of this research involves comparing performance parameters of previous works that developed portable hardware for measuring turbidity from a sample. The turbidimeter presented in \cite{RomanHerrera2016} has a large number of similarities, such as using the arduino platform, a 3D-printed measurement chamber,and a blood test tube for holding samples. This work also compares the results with commercial turbidimeter model HACH® 2100P for calibration of the device. The differences come from the fact that the sensors (emitters and receivers) were different, the circuits for electrical integration of the sensors with the processing unit were as well not the same. Also, in the calibration phase, the work presented here used significantly more data points. Most importantly, the results obtained in this work displayed less error deviation, and or design has a wider turbidity reading range (100 to 1000 NTU versus 50 to 650 NTU).

In the reference \cite{kelley2014affordable} it is achieved a turbidity measuring range of 0 to 1000 NTU, with the lower limit of calibration being analysed at 0.02 NTU, with low standard deviation, indicating that their design is suitable for drinking water quality assessment. The work done by Lambrou, Panayiotou, and Anastasiou \cite{lambrou2014low} is as well an important contribution in the field research for providing a device that measures other significant variables (such as temperature and pH) as well inline real time monitor for drinking water monitoring. The device introduced in this paper is at a disadvantage in the turbidity range of operation, but it can be argued that our work achieved a remarkable accuracy and precision in the range of 100 to 1000 NTU, and that the deviced presented here has its usefulness, when applied to process, food, and pharmaceutical industries, even to government environmental.

The last point to be evaluated is the cost of the prototype, given the objective was the development of a low-cost portable turbidimeter. With respect to the materials used, acquired in the Brazilian market, the developed turbidimeter cost around US\$ $50.00$. Commercial portable turbidimeters, which operate with samples up to $1000$ NTU, have prices in the Brazilian market starting around US\$ $650.00$.

If it would be produced for commercialization, costs of labor and intellectual property would be divided by a certain minimum quantity of devices to be produced. It is possible to hypothesize, even if these costs are estimated, including other factors such as profit margins, the final turbidimeter price would be considerably lower comparing with prices of commercial turbidimeters available on the market. In this analysis, it is reasonable to consider the developed device as being a low-cost portable turbidimeter.

Comparing with other devices, with similar objectives, found in the literature (mainly articles) it is clear that the turbidimeter shown here achieved a good price range. The cheapest device reported is the one in \cite{kelley2014affordable}. They used a microcontroller instead of the whole arduino board, achieving a total cost per prototype of US\$ $35$, and it also stated that scale up could decrease the cost to around US\$ $25$. This approach and cost reduction could also be achieved to our design, in the scale up phase. Reference \cite{Gillett2019}, using similar components and objectives as \cite{kelley2014affordable}, but for continuously monitoring, achieve a total cost of US\$ $64$. Comparing with the turbidimeter with the most similarities, the prototype in reference \cite{RomanHerrera2016} was significantly more expansive, the cost of construction was approximately US\$ 150.

Aside smartphone-based devices, which are relatively 	cheaper, some designs can be found in the literature which are more expensive, above US\$ $70$, due to different goals of the device, such as precise laboratory measurements and educational purposes \cite{Kovacic2019}, in line automatic monitoring \cite{kirkey2018low}, or biomass concentration \cite{nguyen2018low}. There is another published work reporting, and we agree with the authors, that their design achieved low cost, but don't mention the cost achieved by their prototype, which can indicate the cost-range for the commercialized turbidimeter.\cite{metzger2018low}.

Our prototype design, in addition to being a low cost turbidimeter, being also distinguished as portable and allowing serial communication with a computer via a USB port, makes it innovative. Furthermore, the features present in the device developed in this work is hardly found in any commercial solution available in the Brazilian market, especially at low cost.

\section{Conclusions}
\label{sec4}

As proposed, this work dealt with the development of a low-cost portable turbidimeter. The turbidimeter was able to work with turbidity samples in the range of 100 to 1000 NTU, presenting good accuracy and repeatability of the readings within this range. Above 1000 NTU it was not possible to evaluate the effectiveness of the developed turbidimeter, given that the commercial turbidimeter used in the calibration of the developed device is not able to read samples with turbidity greater than that value. Below 100 NTU turbidity readings are not reliable, since the linear model cannot represent the data in this range satisfactorily.

The developed turbidimeter presented material cost around US\$ $ 50.00$. Faced with the prices of commercial turbidimeters, which are found on the market starting around US\$ $650.00$, it is plausible to consider the hardware developed as a low-cost portable turbidimeter. The developed device is still innovative, whereas it has the features of being portable, making use of battery as a power source, and offers to the user the communication with a computer through a USB port, resources difficult to find on the commercial solutions available on the market.

\section{References}
\bibliographystyle{elsarticle-num} 
\bibliography{refsperandio.bib}

\end{document}